\pdfoutput=1
\documentclass[preprint]{elsarticle}
\usepackage{amsmath}
\usepackage{bm}
\usepackage{amssymb,stmaryrd}
\usepackage{mathtools}
\usepackage{algorithmic}
\usepackage{array}
\usepackage{cases}
\usepackage{mdwmath}
\usepackage{mdwtab}
\usepackage{eqparbox}
\usepackage{fixltx2e}
\usepackage{todonotes}
\usepackage{xspace}
\usepackage[vlined,linesnumberedhidden]{algorithm2e}
\usepackage[utf8]{inputenc}
\usepackage[T1]{fontenc}
\usepackage{lmodern}
\usepackage[b]{esvect}
\usepackage{multicol}
\usepackage{graphicx}
\usepackage{subfig}
\usepackage{listings}
\usepackage{xparse}
\usepackage{tabularx,booktabs}
\usepackage{amsthm}
\usepackage{multirow}
\usepackage{varioref}
\usepackage{siunitx}

\newcolumntype{Y}{>{\centering\arraybackslash}X}

\title
{
  Efficient Parallel Solution of the 3D Stationary Boltzmann Transport
  Equation for Diffusive Problems
}

\newcommand{\ndir}{N_{\text{dir}}\xspace}

\newcommand{\pdsa}{PDSA\xspace}

\newcommand{\athos}{{\sc athos}\xspace}

\newcommand{\parsec}{{\sc PaRSEC}\xspace}            % "parsec"
\newcommand{\tbb}{{\sc Intel TBB}\xspace}

\newcommand{\domino}{\textsc{Domino}\xspace}
\newcommand{\diabolo}{\textsc{Diabolo}\xspace}

\newcommand{\dragon}{\textsc{DRAGON}\xspace}

\newcommand{\macrocell}{\texttt{MacroCell}\xspace}   % MacroCell
\newcommand{\macrocells}{\texttt{MacroCell}s\xspace} % MacroCell

\newcommand{\INT}[1]{\llbracket #1 \rrbracket}

\newcommand{\keff}{k_{\text{eff}}}
\newcommand{\normal}[1]{\mathbf{n}_{#1}}
\newcommand{\nouter}{N_{\text{outer}}}

\newcommand{\tspn}{{T_{\text{spn}}}\xspace}
\newcommand{\tsweep}{{T_{\text{sweep}}}\xspace}
\newcommand{\tdomino}{{T_{\text{total}}}\xspace}
\newcommand{\ttotal}{{T_{\text{total}}}\xspace}

\newcommand{\pn}{{\sc P_{\text{N}}}\xspace}
\newcommand{\spn}{{\sc SP_{\text{N}}}\xspace}
\newcommand{\sn}{{\sc S_{\text{N}}}\xspace}

\NewDocumentCommand
\computephi{m m o}{
  \IfNoValueTF{#3}
  {
    ComputePhi(#1,#2)
  }
  {
    ComputePhi(#1,#2,#3)
  }
}

\newcommand{\ncores}{N_{\text{cores}}}

\NewDocumentCommand
\node{m m o}{
  \IfNoValueTF{#3}
  {\mathcal{N}(#1,#2)}
  {\mathcal{N}(#1,#2,#3)}
}

\NewDocumentCommand
\task{o m m o}{
  \IfNoValueTF{#1}
  {
    \IfNoValueTF{#4}
    {\mathcal{T}^{q,#2,#3}}
    {\mathcal{T}^{o,#2,#3,#4}}
  }
  {
    \IfNoValueTF{#4}
    {\mathcal{T}^{#1,#2,#3}}
    {\mathcal{T}^{#1,#2,#3,#4}}
  }
}

\NewDocumentCommand
\taskson{m m o}{
  \IfNoValueTF{#3}
  {\mathcal{T}(#1,#2)}
  {\mathcal{T}(#1,#2,#3)}
}

\NewDocumentCommand
\tasks{o}{
  \IfNoValueTF{#1}
  {\mathcal{U}}
  {\mathcal{U}_f}
}

\newcommand{\tthickrule}{\toprule[2pt]}

\newcommand{\ngroups}{{N_{\text{G}}}\xspace}
\newcommand{\ndofs}{{N_{\text{dof}}}\xspace}

\newcommand{\tcommaccel}{T_{\text{\tiny{PDSA}}}^{\text{comm}}}
\newcommand{\taccel}{T_{\text{\tiny{Accel}}}}

\newtheorem*{iremark*}{Important remark}

\author[edf,inria]{Salli~Moustafa}
\ead{salli.moustafa@edf.fr}
\author[edf]{François~Févotte}
\ead{francois.fevotte@edf.fr}
\author[inria,binp]{Mathieu~Faverge}
\ead{mathieu.faverge@inria.fr}
\author[edf]{Laurent~Plagne}
\ead{laurent.plagne@edf.fr}
\author[inria,ubdx]{Pierre~Ramet}
\ead{pierre.ramet@inria.fr}

\address[edf]{EDF Lab Paris-Saclay - 7, Boulevard Gaspard Monge
 91120 Palaiseau France}
\address[inria]{INRIA Bordeaux - Sud-Ouest, LaBRI, Talence, France}
\address[binp]{Bordeaux INP, Talence, France}
\address[ubdx]{University of Bordeaux, Talence, France}

\begin{document}

\begin{abstract}
This paper presents an efficient parallel method for the deterministic
solution of the 3D stationary Boltzmann transport equation applied to diffusive problems such as nuclear core criticality computations. Based on standard MultiGroup-Sn-DD discretization schemes, our approach combines a highly efficient nested parallelization strategy~\cite{Moustafa2015} with the PDSA parallel acceleration technique~\cite{Fevotte2015} applied for the first time to 3D transport problems. These two key ingredients enable us to solve extremely large neutronic problems involving up to $10^{12}$ degrees of freedom in less than an hour using 64 super-computer nodes.
\end{abstract}

\maketitle

% Message principal à transmettre dans ce papier: on sait modéliser et simuler
% le coeur (l'état neutronique) d'une centrale nucléaire de type REP de façon
% précise et efficace.

\section{Introduction}
\label{sec:introduction}

% Ouvrir la présentation de l'algorithme du Sweep à d'autres disciplines

%What is the problem, why is it hard, why is it important.

This paper presents an efficient parallel deterministic solution of the stationary
Boltzmann Transport Equation (BTE) applied to 3D diffusive problems.

\subsection{Deterministic 3D stationary Boltzmann transport equation solver}
The BTE governs the statistical evolution of gas-like
collections of neutral particles described by phase-space densities $f(\vec{r},\vec{p},t)$ proportional to the number of
particles at a position $\vec{r}$,  with  momentum $\vec{p}$ at  a
given time $t$. This one-body description is widely used to simulate
the transport of particles like neutrons or photons through inhomogeneous reactive
media. The material properties of the media are  characterized by cross-sections that measure the probability of various
particle interactions: absorption, diffusion, emission, etc.

Lying in a six-dimensional (6D) space (3 for space and 3 for
momentum), a precise mesh-based discretization of the stationary BTE
solutions $f(\vec{r},\vec{p})$ can be very large for {\em true} 3D cases where
the considered physical problem offers no particular spatial
symmetry. As an example, using a Cartesian 6D Mesh with 100 points per
axis leads to $10^{12}$ phase-space cells which contain several Degrees Of Freedom (DOFs).
Considering  the scale of a  BTE  solution,  one  can  easily
infer  that  its  solving  procedure  may  rapidly  exhaust  the
capability  of  the  largest  supercomputers. As a consequence, probabilistic methods (Monte-Carlo) that 
avoid the phase-space mesh problems, were the only
approaches  able  to  deal  with  true 3D  cases  until  the  beginning
of  this  century. Unfortunately, probabilistic methods converge slowly with the number $N$ of
pseudo-particles ($\propto N^{-1/2}$) and the computational demand increases strongly with
the desired accuracy. During the last two decades the peak performance
of super-computers has been multiplied by a factor of $10^4$. 
Modern supercomputer capabilities have made deterministic methods a
credible alternative to probabilistic methods for 3D problems and has
allowed for unprecedented accuracy levels for BTE approximate solutions. 

\subsection{Reference criticality computations for nuclear diffusive problems}

BTE solvers are used in different physical contexts and optimal
numerical methods differ from one application to another. In this
paper we address the specific problem of solving the stationary BTE in
diffusive media. Diffusive problems arise when the mean-free
path of particles becomes small compared to the characteristic scale of the considered
problem. For such media, and considering an optically thick enough
geometry, one may neglect the advective part of the transport and
replace the original BTE by the much simpler diffusion equation.

This work takes place in the context of nuclear reactor simulations.
We consider the transport of neutrons inside nuclear reactor cores which
contain optically thick diffusive media. More
specifically, we address the problem of nuclear core criticality
computations. Because nuclear cross-sections mainly depend on the particle
energy, the phase-space density variable $f(\vec{r},\vec{p})$ is replaced
by the angular neutron flux $\psi(\vec{r},E,\vec{\Omega)}=v
f(\vec{r},\vec{p})$ where $\vec{\Omega}$ stands  for  the  particle
momentum direction, $v$ its velocity and $E$ its kinetic
energy. Nuclear operators need to complete many criticality
computations that correspond to stationary BTE solutions.
Industrial routine computations, which are primarily used to conduct operational
and safety studies and to optimize nuclear reactor core
designs, are often based on the diffusion equation
approximation.  In order to assess this approximation, the solution of the
original BTE problem is required. More generally, nuclear operators need
accurate reference transport solutions in order to control the accuracy of their
simulations.

\subsection{Starting from a classical numerical scheme}

The proposed method is based on nested algorithms classically used for nuclear criticality computations. The external loop is a Chebyshev accelerated Power Iteration (PI) that solves the eigenvalue problem $H\psi=k^{-1}F\psi$ where $H$ is the transport operator and $F$ the fission operator. The kinetic energy of neutrons is discretized in well chosen slices called energy groups and, for each PI iteration, an iterative Gauss-Seidel (GS) algorithm is used to solve the multigroup linear problem. For each energy group, the angular variable is treated with the discrete ordinates method ($S_N$) and a Source Iteration (SI) algorithm deals with the coupling between angular components of the flux. %Note that the well known {\em ray effect} limitation of the $S_N$ method
%is not relevant for core simulations since it does
%not affect the quality of the solution in the context of diffusive
%problems.
For diffusive problems the SI procedure converges slowly and  is classically accelerated by the Diffusion Synthetic Acceleration method (DSA)~\cite{Alcouffe1977}. In this
paper, we introduce a parallel extension of the DSA method (\pdsa)
where an efficient single-domain diffusion solver is
required. Finally, the space is discretized over 3D Cartesian meshes,
and all the examples of the paper use the lowest order Diamond
Differencing spatial discretization scheme (DD0), which appears to be
efficient for diffusive nuclear core simulations
\cite{hebert2009}. Note that the \pdsa method does not depend on the
DD0 choice and a higher order numerical scheme could have been used. The only condition is that these alternative schemes must be consistent with the single-domain DSA solver.

\subsection{New metrics for efficient numerical algorithms}
The tremendous peak power of modern supercomputers, that commonly
exceeds $10^{16}$ floating point operations per second (FLOPS), is
accompanied by a high architecture complexity. Indeed, recent architectures
exhibit a hierarchical organization (cluster of nodes of multicore
processors with vector units) which requires a mix
of different parallel programming paradigms (Message Passing,
Multi-Threading, SIMD) to achieve optimal efficiency. In addition to this mixture of
parallel programming models, a new constraint on the data movements has
emerged and plays a dominant role in computation efficiency. A direct
consequence of this machine evolution is that numerical algorithms
should no longer be evaluated upon their {\em parallel scalability} ($i$)
alone. The {\em computational density} ($ii$) which measures the
ratio between the number of floating point (FP) operations and the number
of data movements from the off-chip memory to CPU registers is a new metric that
must be considered to evaluate the efficiency of a given
algorithm. Finally,  the {\em vectorization potential} ($iii$) of a given
algorithm will determine its ability to benefit from the ever
increasing Single Instruction Multiple Data (SIMD) width of dedicated
CPU FP SIMD instructions (SSE2, AVX, AVX512). The
combination of these three algorithm characteristics ($i$,$ii$,$iii$)
will eventually result in an efficient
numerical solver. In this paper we put a particular focus on the
efficiency of the proposed BTE solver and explain how {\em parallel
  scalability}, {\em computational density} and {\em vectorization
  potential} are taken into account.

\subsection{Paper contributions}

In this paper we propose a parallel and efficient solution method for
the stationary BTE that allows one to carry out very large
criticality computations for diffusive problems on moderately large
supercomputers. These {\em affordable} full 3D
transport computations result from an uncommonly high effective FP
performance that can exceed 20\% of the available theoretical peak performance of
the computing nodes. As a representative example, we show that a 3D PWR $\keff$ computation with 26 energy
groups, 288 angular directions, and $578\times578\times140$ space
cells, can be completed in less than an hour using 64 cluster
nodes. Two main ingredients are combined in the proposed method that has been implemented in \domino~\cite{dominoMC13,Moustafa2015ANE}, our in-house neutron transport solver. 
\begin{itemize}
  \item A very high performance {\bf sweep} algorithm including 3 nested levels of
    parallelism with good data locality and fine grained synchronization
    that has been described in detail in~\cite{Moustafa2015}.
  \item A novel scalable \pdsa acceleration technique for diffusive problem
    introduced in~\cite{Fevotte2015} and applied for the first time to 3D transport
    computations. This method is easy to implement provided a fast
    single-domain shared-memory diffusion solver. Hence \pdsa allows
    one to avoid the complex task of building fully distributed diffusion
    solvers as implemented in \cite{Jamelot2013,Lathuiliere2011}.
\end{itemize}

Recently, important progress has been made for increasing the
scalability of BTE solvers. In \cite{denovo} the
authors replace the Power Iterations by advanced eigenvalue algorithms and treat the energy groups in
parallel. The scalability of this approach is impressive and parallel
computations involving more than $10^5$ computing cores are presented.
In this current paper we show that, for a moderately high number of groups
($\le$ 26), the proposed method results in fast criticality
computations with more modest numbers of computing cores ($10^2-10^3$)
thereby making 3D stationary computation more affordable.

This result should have an impact on the acceleration strategies for
other kinds of BTE solvers like unstructured mesh based transport
solvers or the accelerated Monte-Carlo approach for criticality nuclear computations.
%\ref{Denovo2}, author adress the problem of performing BTE
%computations on GPUs.
%%Other works focus on reducing the number of DOFs requied to reach a given accuracy. Unstructured
%%spatial meshes, adaptive mesh refinement, wavelet method for angles,\dots.
\\

The paper is organized as follows. In Section~2, we describe  the equations to be solved,
the different discretization schemes, the main algorithm and the three
nested levels of parallelism used in the sweep implementation described in~\cite{Moustafa2015}. 
In Section~3, the \pdsa algorithm and its
implementation are described and some details are given regarding the correct
coupling between the Transport DD0 discretization and the Finite
Element method used in the \pdsa. Section~4 describes the
parallel performance achieved by \domino for different PWR criticality computation configurations. Some
conclusions and outlooks are given in section~5.

\section{The Discrete Ordinates Method for Neutron Transport
  Simulation}
\label{sec:sn_method}
% \begin{itemize}
% \item quadrature formula
% \item source iterations
% \item sweep
% \end{itemize}
\subsection{Source Iterations Scheme}
\label{subsec:source_iterations}
We consider the monogroup transport equation as defined in
equation~\eqref{eq:monogroup_equation}.
\begin{align}
  \label{eq:monogroup_equation}
  \begin{split}
    \underbrace
    {
      \vv \Omega \cdot \vv \nabla\psi(\vv r, \vv \Omega)
      +
      \Sigma_t(\vv r, \vv \Omega)
      \psi(\vv r, \vv \Omega)
    }_{L\psi(\vv r, \vv \Omega)}
    -
    \overbrace
    {
      \int_{S_2}{d\vv \Omega'}
      \Sigma_s(\vv r, \vv \Omega' \cdot \vv \Omega)
      \psi(\vv r, \vv \Omega')
    }^{R(\vv r, \vv \Omega)}
    =
    Q(\vv r, \vv \Omega),
  \end{split}
\end{align}
where $Q(\vv r, \vv \Omega)$ gathers both monogroup fission and
inter-group scattering sources. The angular dependency of this
equation is resolved by looking for solutions on a discrete set of
carefully selected angular directions \{$\vv \Omega_i \in S_2$,
$i=1,2,\cdots, \ndir$\}, called discrete ordinates; each one being
associated to a weight $w_j$. In general, the discrete ordinates are
determined thanks to a numerical quadrature formula as defined in
equation~\eqref{eq:quadrature}. For any summable function $g$ over $S_2$:
\begin{equation}
  \label{eq:quadrature}
  \bar g
  \equiv
  \int_{S_2}g(\vv\Omega)d\vv\Omega
  \simeq
  \sum_{j=1}^{\ndir} w_jg(\vv\Omega_j).
\end{equation}
In \domino, we use the \textit{Level Symmetric}~\cite{carlsonlee}
quadrature formula, which leads to $\ndir=N(N+2)$ angular directions,
where $N$ stands for the \textit{Level Symmetric} quadrature formula
order.

Therefore, considering that the cross-sections are isotropic,
equation~\eqref{eq:monogroup_equation} becomes:
\begin{align}
  \label{eq:monogroup_equation_discrete}
  \begin{split}
    \vv \Omega_i \cdot \vv \nabla\psi(\vv r, \vv \Omega_i)
    +
    \Sigma_t(\vv r)
    \psi(\vv r, \vv \Omega_i)
    =
    \overbrace
    {
      Q(\vv r, \vv \Omega_i)
      +
      \Sigma_s(\vv r)
      \phi(\vv r)
    }^{S(\vv r, \vv \Omega_i)},
  \end{split}
\end{align}
where $\phi(\vv r)$ is the scalar flux and defined by:
\begin{equation}
  \label{eq:scalar-flux}
  \phi(\vv r) \equiv \bar\psi(\vv r, \cdot)
  =
  \int_{S_2}\psi(\vv r, \vv\Omega)d\vv\Omega
  \simeq
  \sum_{j=1}^{\ndir} w_j\psi(\vv r, \vv\Omega_j).
\end{equation}
Equation~\eqref{eq:monogroup_equation_discrete} is solved by iterating
over the scattering source as described in
Algorithm~\ref{alg:scattering}.
\begin{algorithm}[!htbp]
  \caption{Source iterations}
  \label{alg:scattering}
  \SetKwInOut{Input}{Input}
  \SetKwInOut{Output}{Output}
  \Input{$\phi^k$}
  \Output{$\phi^{k+\frac{1}{2}}$}
  \While{Non convergence}
  {
    \For{$i=1, \dots, \ndir$}
    {
      $
      S(\vv r, \vv \Omega_i)
      =
      Q(\vv r, \vv \Omega_i)
      +
      \Sigma_s(\vv r)\phi^k(\vv r)
      $\;
      \strut
      \ShowLn
      $
      L\psi^{k+\frac{1}{2}}(\vv r, \vv \Omega_i)
      =
      S(\vv r, \vv \Omega_i)
      $\;\label{ln:fixed_source}
      $
      \phi^{k+\frac{1}{2}}(\vv r)
      =
      \sum_{j=1}^{\ndir} w_j\psi^{k+\frac{1}{2}}(\vv r, \vv\Omega_j)
      $\;
    }
  }
\end{algorithm}

Each source iteration (SI) involves the
resolution of a fixed-source problem (Line~\ref{ln:fixed_source}), for
every angular direction. This is done by discretizing the spatial
variable $\vv r$ of the streaming operator $L$. In this work, we focus
on a $3$D reactor core model, represented by a $3$D Cartesian domain
$\mathcal{D}$, and $L$ is discretized using a Diamond Difference
scheme (\textsc{DD}), as presented by A. Hébert
in~\cite{hebert2006}. The discrete form of the fixed-source problem is
then solved by ``walking'' step by step throughout the whole spatial
domain and to progressively compute angular fluxes in the spatial
cells. In the literature, this process is known as the \textbf{sweep
operation}. The vast majority of computations performed in
the $\sn$ method are part of the sweep operation.

\subsection{Sweep Operation}
\label{subsec:sweep}
\begin{figure*}[!htbp]
  \centering
  \subfloat
  [Angular quadrature in 2D. Directions are grouped in quadrants.]
  {
    \label{fig:quadrature}
    \includegraphics[width=0.2\linewidth]{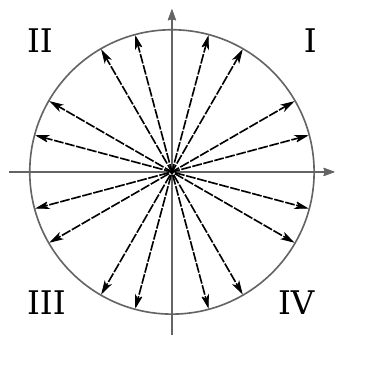}
  }
  \qquad
  \subfloat
  [
  In each direction, cells have two incoming components of the flux
  (Here, from the left and bottom faces: $\psi_L$ and $\psi_B$), and
  generates two outgoing components of the flux (Here, on the right and
  top faces: $\psi_R$ and $\psi_T$).
  ]
  {
    \label{fig:cell-2D}
    \includegraphics[width=0.3\linewidth]{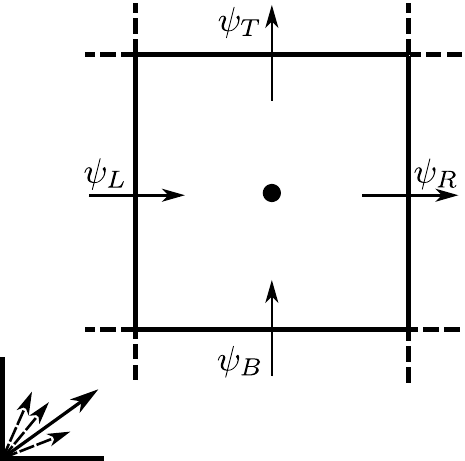}
  }
  \qquad
  \subfloat
  [
  Domain decomposition and boundary conditions. The corner cell
  $(0,0)$ is the first to be processed for a quadrant, and its
  processing then allows for the processing of its neighbors (Here $(0,1)$ and
  $(1,0)$).
  ]
  {
    \label{fig:sweep-2D-6X6-grid}
    \includegraphics[width=0.3\linewidth]{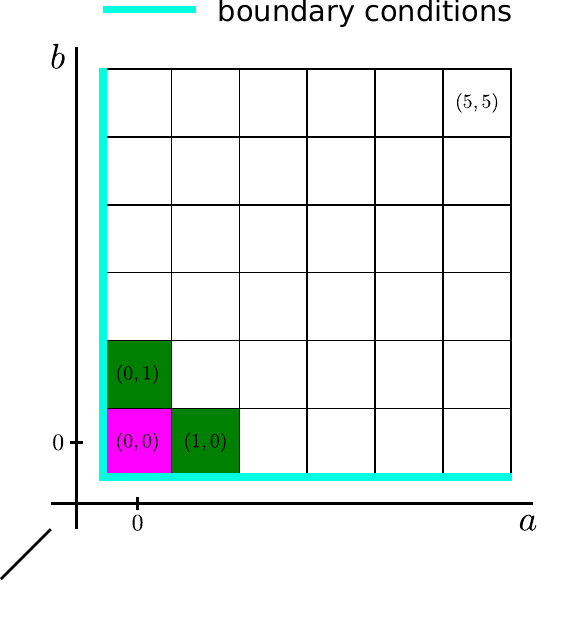}
  }
  \caption
  {
    Illustration of the sweep operation over a $6 \times 6$ 2D spatial
    grid for a single direction.
  }
  \label{fig:sweep-2D-6x6}
\end{figure*}

The sweep operation is used to solve the space-angle problem on
Line~\ref{ln:fixed_source} of Algorithm~\ref{alg:scattering}. It
computes the angular neutron flux inside all cells of the spatial
domain, for a set of angular directions. These directions are grouped
into four quadrants in 2D (or eight octants in 3D). In the following,
we focus on the first quadrant (labeled I in
Figure~\ref{fig:quadrature}). As shown in Figure~\ref{fig:cell-2D} explanation,
each cell has two incoming dependencies $\psi_L$ and $\psi_B$ for each
angular direction. At the beginning, incoming fluxes on all left and
bottom faces are known as indicated in
Figure~\ref{fig:sweep-2D-6X6-grid}. Hence, the cell $(0,0)$ located at
the bottom-left corner is the first to be processed. The treatment of
this cell allows for the updating of outgoing fluxes $\psi_R$ and
$\psi_T$, that satisfy the dependencies of cells $(0,1)$ and
$(1,0)$. These dependencies on the processing of cells define a
sequential nature throughout the progression of the sweep operation:
two adjacent cells belonging to successive diagonals cannot be
processed simultaneously. However, all cells belonging to a same diagonal can be
processed in parallel. Furthermore, treatment of a single cell for
all directions of the same quadrant can be done in
parallel. Hence, step by step, fluxes are evaluated in
all cells of the spatial domain, for all angular directions belonging
to the same quadrant. The same operation is repeated for all the four
quadrants. When using vacuum boundary conditions, there are no incoming
neutrons to the computational domain and therefore processing of the
four quadrants can be done concurrently. This sweep operation is
subject to numerous studies regarding design and parallelism to reach
highest efficiency on parallel architectures.% In
%section~\ref{sec:hierarchical_sweep}, we will present our task-based
%approach that enables us to leverage the full computing power of such
%architectures.

\subsection{Hierarchical Parallelization of the Sweep}
%\section{Massively Parallel Implementation of the $\sn$ Method}
\label{sec:hierarchical_sweep}
In this section, we briefly describe the parallelization of the $\sn$-sweep
operation on distributed multicore-based architectures. A detailed description of
the \domino's $\sn$-sweep can be found in \cite{Moustafa2015} and \cite{thesesalli}.

\begin{figure*}[!htbp]
  \centering
  \subfloat
  [A 2D single-quadrant Sweep's DAG over a 4x4 grid of \macrocells.]
  {
    \label{fig:sweep-2D-4X4-dag}
    \includegraphics[width=0.35\linewidth,height=0.45\linewidth]{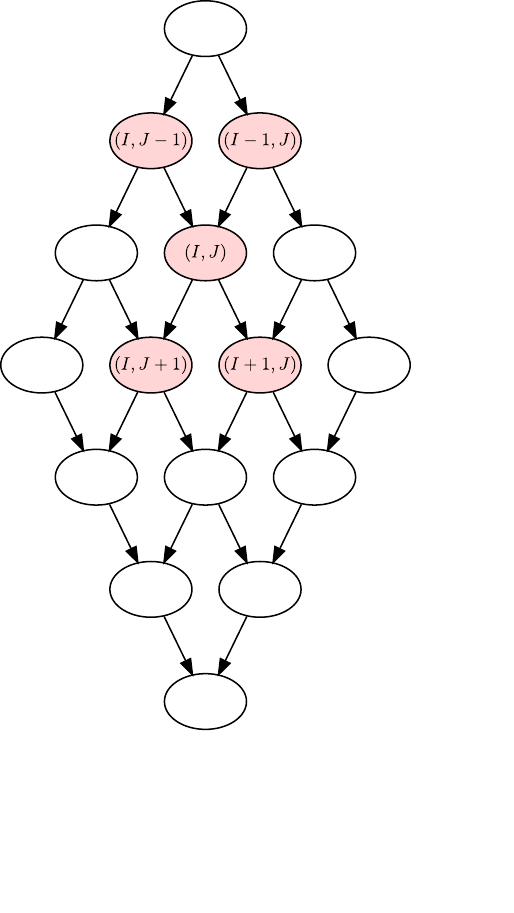}
  }
  \qquad
  \subfloat
  [
  Snapshot of an execution of the Sweep operation implemented on
    top of \parsec. \macrocells of similar colors are processed on the
    same node and highlighted ones are those that are in the process of being
    executed (at the time the snapshot was taken).
  ]
  {
    \label{fig:r480M20N11P2Q2R2_PBQ_Front}
    \includegraphics[width=0.5\linewidth]{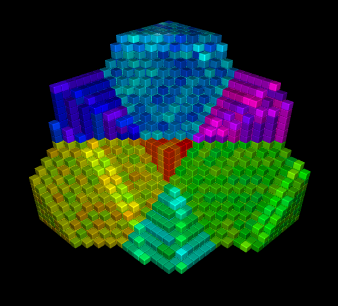}
  }
  \caption
  {
    Sweep's DAG and snapshot.
  }
  \label{fig:dagandsnapshot}
\end{figure*}

As one can see from Figure~\ref{fig:sweep-2D-6x6}, a space cell $c_{i,j}$ with Cartesian indices $i$ and $j$ can be processed as soon
as both cells  $c_{i-1,j}$ and $c_{i,j-1}$ have been computed. In order to reduce the cost of parallel communications, we do not consider individual cells but group them into \macrocells $C_{I,J}$ that correspond to rectangular sets of cells. Let $T_{I,J}$ be the task corresponding to the sweep inside a \macrocell $C_{I,J}$. The dependency between all the tasks:
\[
(T_{I-1,J},T_{I,J-1})\rightarrow T_{I,J}
\]
defines a Directed Acyclic Graph (DAG) which corresponds to the
complete sweep from one corner of the spatial mesh to the opposite
corner. An illustration of this DAG is presented in
Figure~\ref{fig:sweep-2D-4X4-dag}.

%\begin{figure}[!htpb]
%  \centering
%  \includegraphics[width=0.5\linewidth]{figures/svg/sweep-2D-4X4-dag_Full}
%  \caption
%  {
%    A 2D single-quadrant Sweep's DAG over a 4x4 grid of \macrocells.
%  }
%  \label{fig:sweep-2D-4X4-dag}
%\end{figure}
%% This DAG description, the task implementation and the data
%% distribution over the computing nodes, are passed to the \parsec
%% parallel framework. \parsec generates, at compile time, a two-level
%% parallel implementation. At the outer level, the communications
%% between the computing nodes are handled via a generated MPI
%% implementation. At the inner level, the communications between all
%% cores inside each node are handled via a generated multi-threaded
%% implementation. The \parsec runtime schedules the execution of the
%% defined sweep DAG for all the nodes and all the cores. A snapshot of
%% such an execution is depicted in Figure~\ref{fig:r480M20N11P2Q2R2_PBQ_Front}.
This DAG description, the task implementation and the data
distribution over the computing nodes, are passed to a parallel
runtime system. Here, we choose the \parsec~\cite{Bosilca201237} runtime system and
its specific parameterized task graph to describe the algorithm. This
format corresponds well with the regular pattern of our regular domain
decomposition and allows the runtime system to schedule the tasks in
a fully distributed manner without discovering integrally the graph of
dependencies. In practice, \parsec exploits this pattern regularity to automatically schedule all
computations on a set of threads per node (usually one thread per core), and
triggers communications through an MPI layer when necessary.
A snapshot of the execution on top of \parsec is depicted in
Figure~\ref{fig:r480M20N11P2Q2R2_PBQ_Front}.

%% At the outer level, the communications
%% between the computing nodes are handled via a generated MPI
%% implementation. At the inner level, the communications between all
%% cores inside each node are handled via a generated multi-threaded
%% implementation. The \parsec runtime schedules the execution of the
%% defined sweep DAG for all the nodes and all the cores. A snapshot of
%% such an execution is depicted in Figure~\ref{fig:r480M20N11P2Q2R2_PBQ_Front}.
%\begin{figure}[!htpb]
%  \centering
%  \includegraphics[width=0.5\linewidth]{figures/png/r480M20N11P2Q2R2_PBQ_Front.png}
%  \caption
%  {
%    Snapshot of an execution of the Sweep operation implemented on
%    top of \parsec. \macrocells of similar colors are processed on the
%    same node and highlighted ones are those that were being
%    executed.
%  }
%  \label{fig:r480M20N11P2Q2R2_PBQ_Front}
%\end{figure}
In the case of vacuum boundary conditions, all 4 (resp. 8) angular
sweep quadrants (resp. octants) are processed in parallel and again,
handled via \parsec.

Finally, each task $T_{I,J}$ exhibits a third innermost level of
parallelism based on Single Instruction Multiple Data (SIMD). SIMD
capabilities of modern computing cores allow them to perform, at each
clock cycle, several identical floating point operations
($+$,$*$,\dots) on different floating point values. This SIMD
parallelism is used to perform the sweep operations that correspond to
different angular directions of the same quadrant inside each spatial
cell. A detailed description of this {\em vectorized} implementation
based on Eigen, a C++ template library, can be found in~\cite{Moustafa2015ANE}.

\section{Acceleration of Scattering Iterations using PDSA}
\label{sec:pdsa}
In highly diffusive media ($\Sigma_s\approx\Sigma_t$), the convergence
of Algorithm~\ref{alg:scattering} is very slow, and therefore a
numerical acceleration scheme must be combined with this algorithm in
order to speed-up its convergence. One of the widely used acceleration
schemes in this case, is Diffusion Synthetic Acceleration~(DSA)~\cite{Larsen2010}.

\subsection{Diffusion Synthetic Acceleration}
\label{subsec:dsa}
\begin{figure}[!htpb]
  \centering
  \includegraphics[width=0.2\textwidth]{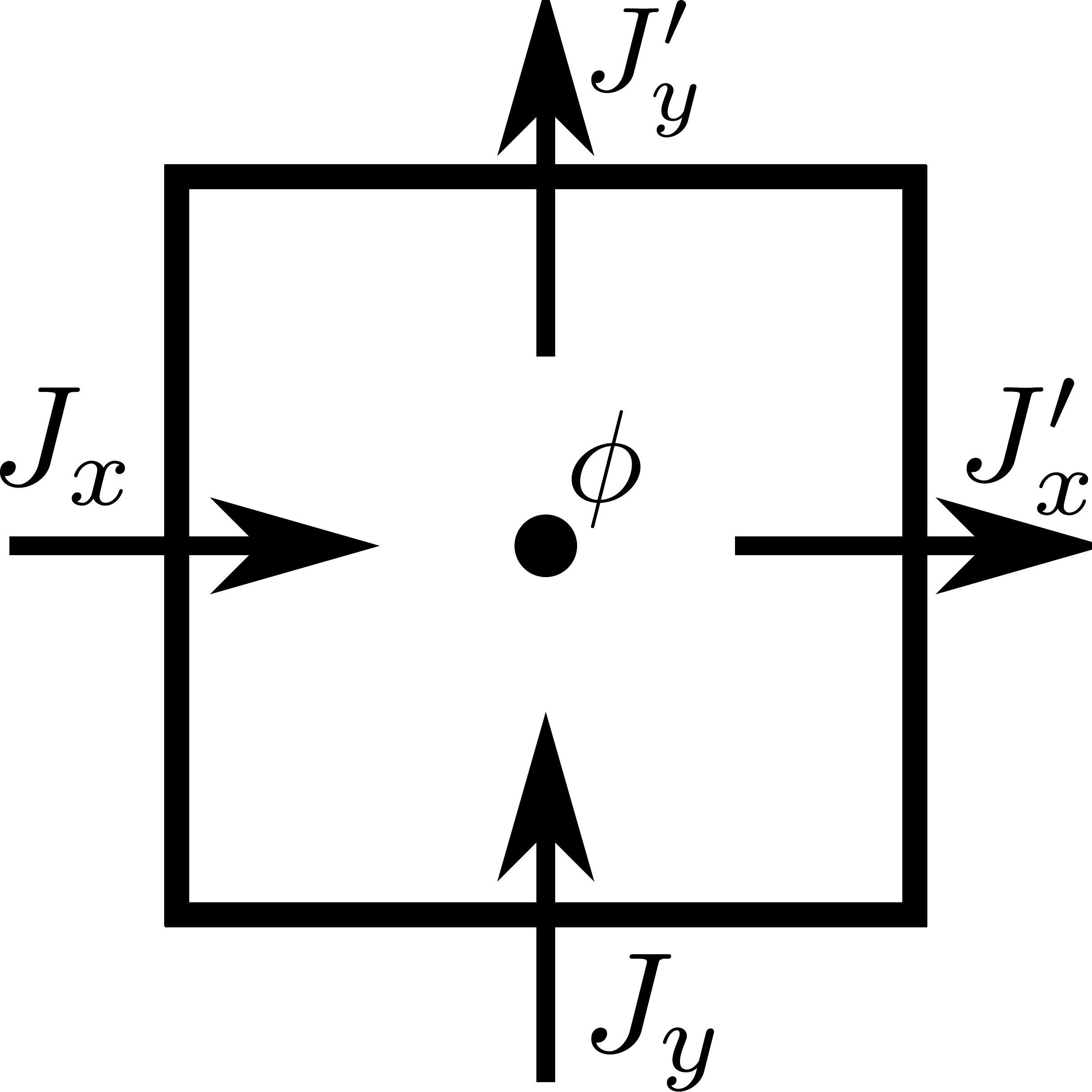}
  \caption
  {
    RT0 finite element in 2D: $5$ DoFs ($4$ for the currents
    $J_x$, $J'_x$, $J_y$, $J'_y$ and $1$ for the scalar flux
    $\phi$). DSA is applied using an RT0 element.
  }
  \label{fig:ddl_rt0}
\end{figure}

Here we just recall the basics of this method, and the reader can
refer to the paper~\cite{Larsen2010} for more details regarding its
effectiveness and the Fourier analysis characterizing its convergence
properties. Let us define
$\epsilon^{k+\frac{1}{2}}=\psi-\psi^{k+\frac{1}{2}}$, as the error of
the solution obtained after the $k+\frac{1}{2}^{\text{th}}$ iteration
of the SI scheme, relative to the exact solution $\psi$, as defined by
equation~\eqref{eq:monogroup_equation_discrete}. The error $\epsilon$
satisfies the following transport equation:
\begin{equation}
  L\epsilon^{k+\frac{1}{2}}(\vv r, \vv\Omega_i)
  =
  \Sigma_t(\vv r)
  \bar\epsilon^{k+\frac{1}{2}}
  +
  \Sigma_s(\vv r)
  \displaystyle
  \left(
    \phi^{k+\frac{1}{2}}(\vv r) - \phi^k(\vv r)
  \right),
  \label{eq:iterate_si}
\end{equation}
where $\bar\epsilon$ is the scalar field associated to $\epsilon$ and
defined as in equation~\eqref{eq:quadrature}. However,
equation~\eqref{eq:iterate_si} is as difficult to solve as the original
fixed-source transport
problem~\eqref{eq:monogroup_equation_discrete}. Nevertheless, if an
approximation $\tilde\epsilon$ of $\bar\epsilon$ was available, then
the scalar flux could be updated to:
\[
  \phi^{k+1}(\vv r)
  =
  \phi^{k+\frac{1}{2}}(\vv r)+\tilde\epsilon^{k+\frac{1}{2}}(\vv r).
\]
The idea of the DSA method is then to use a diffusion approximation,
yielding $\tilde\epsilon$, instead of solving the transport
equation~\eqref{eq:iterate_si}. In \domino, the diffusion
approximation is obtained using the \diabolo solver~\cite{plagnespn},
which implements the simplified~$\pn$~($\spn$) method as presented
in~\cite{Lautard1999}, in a mixed-dual formulation. When approximating
equation~\eqref{eq:iterate_si} with a diffusion operator, the problem
solved by \diabolo can be stated as the following mixed
dual formulation:
\begin{align}
  &\text{Find $(\tilde\epsilon^{k+\frac{1}{2}},\vv j^{k+\frac{1}{2}})
    \in L^2(\mathcal{D}) \times H(\mathcal{D},\operatorname{div})$ such that:}\notag\\
  &\quad\left\{
    \begin{aligned}
      &\operatorname{div}\vv j^{k+\frac{1}{2}}(\vv r)
      +\Sigma_{a}\tilde\epsilon^{k+\frac{1}{2}}(\vv r)
      =   \Sigma_s(\vv r)
      \displaystyle
      \left(
        \phi^{k+\frac{1}{2}}(\vv r) - \phi^k(\vv r)
      \right)
      &\text{in}~\mathcal{D},\\
      &\frac{1}{D}\vv j^{k+\frac{1}{2}}(\vv r)
      + \vv{\nabla}\;\tilde\epsilon^{k+\frac{1}{2}}(\vv r)
      = \vv 0
      &\text{in}~\mathcal{D},\\
      &\tilde\epsilon^{k+\frac{1}{2}}=0   &\text{on}~\partial\mathcal{D},
    \end{aligned}
  \right.
  \label{eq:dsa_sp1}
\end{align}
in which we introduced the diffusion coefficient $D$ and the neutronic
current~$\vv j^{k+\frac{1}{2}}$ associated to
$\tilde\epsilon^{k+\frac{1}{2}}$. Within \diabolo, these equations are spatially
discretized using an RTk finite elements scheme~\cite{Raviart1977,Nedelec1986}
(see Figure~\ref{fig:ddl_rt0}), which is consistent with the DD scheme used for
the discretization of the transport equation as proven
in~\cite{hebert2006}. Therefore, the stability of the
acceleration scheme is ensured.

However, when integrated into a parallelized transport solver, DSA may
become a bottleneck for the scalability of the transport solver if,
for instance, a serial implementation of the diffusion solver is
used. On the other hand, if the diffusion solver is parallelized, as
presented in~\cite{Jamelot2013,Lathuiliere2011} using a domain
decomposition method, the iteration count to the solution increases
with the number of subdomains, and can lead to a poor global
scalability~\cite{Yavuz1992}. To remedy this issue, a variant of the
DSA scheme has been recently proposed by F. Févotte in~\cite{Fevotte2015}.

\subsection{Piecewise Diffusion Synthetic Acceleration}
\label{subsec:pdsa}
The general presentation and the convergence proof of the Piecewise Diffusion
Synthetic Acceleration (\pdsa) method are given in~\cite{Fevotte2015}. We recall
that the purpose of this method is similar to that of DSA: evaluate an
approximation, $\tilde\epsilon^{k+\frac{1}{2}}$, of the error on the scalar flux, $\bar\epsilon^{k+\frac{1}{2}}$, to be used for correcting the scalar flux, $\phi^{k+\frac{1}{2}}$. In the following, iteration indices $k+\frac{1}{2}$ will
be dropped for the sake of readability.

We assume that the spatial domain $\mathcal{D}$ is split, along the
$3$ spatial dimensions, into $N=P \times Q \times R$
non-overlapping subdomains $\mathcal{D}_I$ such that:
$\mathcal{D}=\cup_{I \in \mathcal{I}}\mathcal{D}_I$, where
\begin{equation*}
  \mathcal{I}
  =
  \INT{1,P} \times \INT{1,Q} \times \INT{1,R}.
\end{equation*}
We set: $\Gamma_{IJ}=\partial\mathcal{D}_I\cap\partial\mathcal{D}_J$
the non-empty interfaces between subdomains of index $I$ and $J$;
$\Gamma_I=\partial\mathcal{D}\cap\partial\mathcal{D}_I$ and
$\normal{I}$ the unit normal vector to $\partial\mathcal{D}_I$ and
$\tilde\epsilon^I=\tilde\epsilon|_{\mathcal{D}_I}$ and
$\vv j^I=\vv j|_{\mathcal{D}_I}$,
 the respective restrictions of $\tilde\epsilon$ and $\vv j$ to subdomain~$\mathcal{D}_I$.

Unlike the DSA method which consists in solving a single $\spn$ problem on
the global domain $\mathcal{D}$, the \pdsa method is based on
successive resolutions of two $\spn$ problems on each of the subdomains
$\mathcal{D}_I$. These $\spn$ problems differ on the boundary
conditions applied on the subdomains: the first problem uses
homogeneous Neumann boundary conditions
(equation~\eqref{eq:sp1_symmetry_bc}), and the second one uses
non-homogeneous Dirichlet boundary conditions
(equation~\eqref{eq:sp1_mean_flux}). In both equations, notations were shortened
by using $S^I(\vv r)$ to denote the right-hand side of equation~\eqref{eq:dsa_sp1}.

\begin{equation}
  \left\{
    \begin{aligned}
      &\operatorname{div}\vv j_N^I(\vv r)+\Sigma_{a}\tilde\epsilon_N^I(\vv r)=S^I(\vv r) &\text{in}~\mathcal{D}_I\\
      &\vv{\nabla}\;\tilde\epsilon_N^I(\vv r)+\frac{1}{D}\vv j_N^I(\vv r)=\vv 0 &\text{in}~\mathcal{D}_I\\
      &\tilde\epsilon_N^I=0   &\text{on}~\Gamma_I\\
      &\vv\nabla\tilde\epsilon_N^I \cdot \normal{I}=0 &\text{on}~\Gamma_{IJ}
    \end{aligned}
  \right.
  \label{eq:sp1_symmetry_bc}
\end{equation}
\begin{equation}
  \left\{
    \begin{aligned}
      &\operatorname{div}\vv j_D^I(\vv r)+\Sigma_{a}\tilde\epsilon_D^I(\vv r)=S^I(\vv r) &\text{in}~\mathcal{D}_I\\
      &\vv{\nabla}\;\tilde\epsilon_D^I(\vv r)+\frac{1}{D}\vv j_D^I(\vv r)=\vv 0 &\text{in}~\mathcal{D}_I\\
      &\tilde\epsilon_D^I=0  &\text{on}~\Gamma_I\\
      &\tilde\epsilon_D^I=\frac{\tilde\epsilon_N^I+\tilde\epsilon_N^J}{2}  &\text{on}~\Gamma_{IJ}
    \end{aligned}
  \right.
  \label{eq:sp1_mean_flux}
\end{equation}
As shown in~\cite{Fevotte2015}, for sufficiently diffusive and optically thick
problems, the PDSA solution~$\tilde\epsilon_D$ is an approximation of the global
DSA solution $\tilde\epsilon$.
The accelerated $\sn$ flux is thus finally given by:
\[
  \phi^{k+1}=\phi^{k+\frac{1}{2}}+\tilde\epsilon_D.
\]

\bigskip

\begin{figure*}[!htbp]
  \centering
  \subfloat
  [
  The first step consists of solving two diffusion problems in
  parallel on $\mathcal{D}_1$ and $\mathcal{D}_2$, with Neumann
  boundary conditions.
  ]
  {
    \label{fig:pdsa-neumann}
    \includegraphics[width=0.8\linewidth]{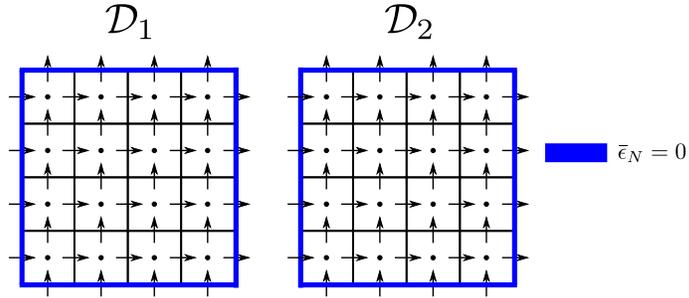}
  }

  \subfloat
  [
  The second step also solves two diffusion problems, but with
  non-homogeneous Dirichlet boundary conditions: null flux boundary
  conditions on the external boundary of the domain and an average
  value of the flux at the inner interface.
  ]
  {
    \label{fig:pdsa-dirichlet}
    \includegraphics[width=0.8\linewidth]{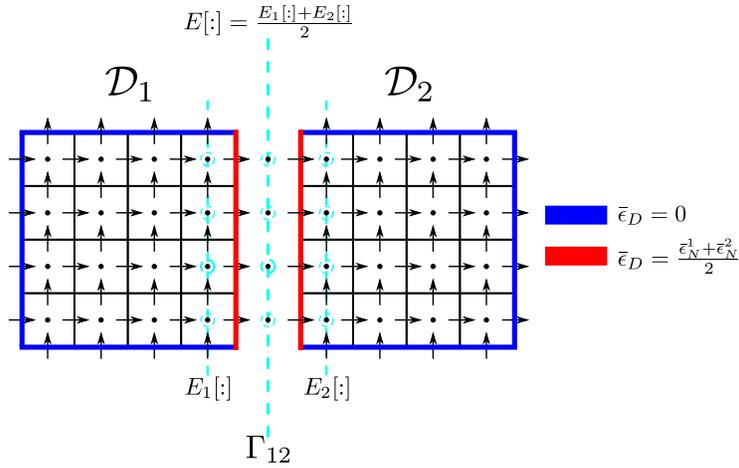}
  }
  \caption
  {
    Illustration of the \pdsa method on a domain split in two.
  }
  \label{fig:pdsa-boundary-conditions}
\end{figure*}

On the practical side, it is important to note that the application of the \pdsa method using
a classical diffusion solver does not require many changes. Homogeneous Dirichlet
boundary conditions used over $\Gamma_I$ are classically implemented to simulate
whole cores; homogeneous Neumann boundary conditions used over $\Gamma_{IJ}$ in
equation~\eqref{eq:sp1_symmetry_bc} are likewise featured by most diffusion
solvers to implement symmetric domains. However, the second PDSA
step~\eqref{eq:sp1_mean_flux} requires implementing a non-homogeneous Dirichlet boundary
condition, which is not standard. In our case we are using a mixed-dual formulation of the
$\spn$ equations, therefore these boundary conditions are natural and not
essential and their implementation is straightforward. An illustration of
the processing of the boundary conditions in the case of two
subdomains is presented in Figure~\ref{fig:pdsa-boundary-conditions}.
This is a major shift from the classical DSA method, as we are no
longer required to get the solution of the diffusion problem on the
whole spatial domain. The first advantage of this method is that the
explicit global synchronizations between the resolutions of the
piecewise diffusion problems are largely reduced, hence allowing to
fully parallelize the DSA method without efficiency loss. In addition,
as we are going to see in section~\ref{sec:experiments}, the
effectiveness of the \pdsa method is comparable to that of the
classical DSA method on a class of benchmarks.

\subsection{Parallelization of the \pdsa Method}
\label{subsec:task_based_sweep}
\begin{figure*}[!htpb]
  \centering
  \includegraphics[width=0.9\linewidth]{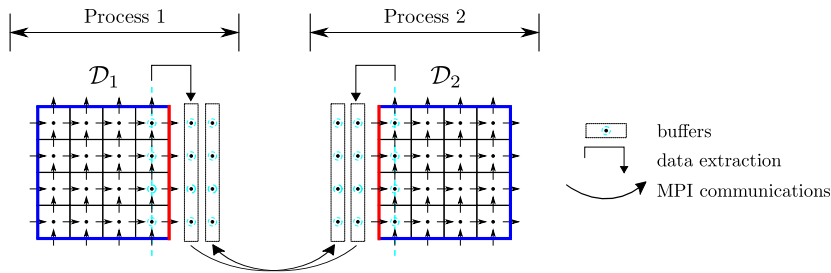}
  \caption
  {
    Illustration of the communication pattern in the \pdsa method
    on a domain split in two. Two point-to-point communications are
    needed to exchange flux information at the interface between the two
    subdomains.
  }
  \label{fig:pdsa_comm}
\end{figure*}

Figure~\ref{fig:pdsa_comm} illustrates a parallel implementation of
the \pdsa method in 2D, when the global domain is partitioned in two
subdomains.
The partitioning of the global domain uses the same block data
distribution as for the sweep operation. As we mentioned previously in
section~\ref{subsec:dsa}, the diffusion problem on each subdomain is
solved using our $\spn$ solver \diabolo which is parallelized on
shared memory systems using the \tbb framework.

Hence, by mapping each subdomain to a single process, the resolution
of the diffusion problems on $\mathcal{D}_1$ and $\mathcal{D}_2$, when
applying the \pdsa method, is naturally performed in parallel. Moreover,
for the first step, the use of Neumann boundary conditions requires no
communication with the neighboring processes. However, in the second
step, each process needs to have the average value of the scalar flux
at the interfaces between its neighbors. Therefore, each process must
perform send and receive operations to exchange data with its
neighbors.
%These data exchanges are point-to-point communications as
%only two processes are involved for each data exchange. To achieve
%this, we allocate two extra buffers \todo{\footnotesize be more allusive: don't mention
%the buffers, since a proper implementation should take advantage of ParSEC for
%this too} per process: the first one is
%dedicated to store the extracted scalar flux at the interface between
%the subdomains which is then communicated to the neighboring process;
%while the second one is used as a reception buffer. We use
%asynchronous MPI communication primitives to exchange the flux at the
%interfaces.

\section{Performance of \domino}
\label{sec:experiments}
The performance and accuracy of the single-domain \domino implementation, based on standard
DSA acceleration, has been assessed in \cite{thesesalli} and \cite{dominoMC13} where
comparisons with both reference Monte-Carlo and deterministic solvers have been conducted.
In this section, we assess the performance of the present multi-domain \domino implementation
based on the \pdsa method for solving a set of PWR nuclear core benchmarks.

\begin{figure}[!htbp]
  \centering
  \subfloat[Material Radial Map]{
    \includegraphics[width=0.3\linewidth]{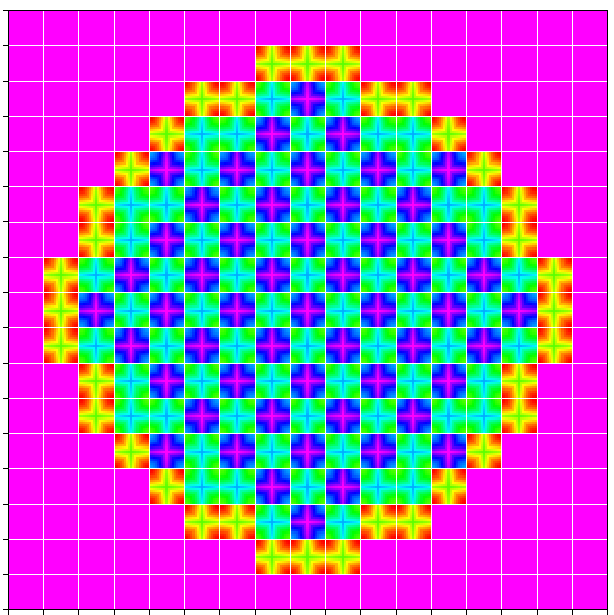}
    \label{fig:PWR900_materials}    
  }
  ~
  \subfloat[Neutron Thermal Flux]{
    \includegraphics[width=0.6\linewidth]{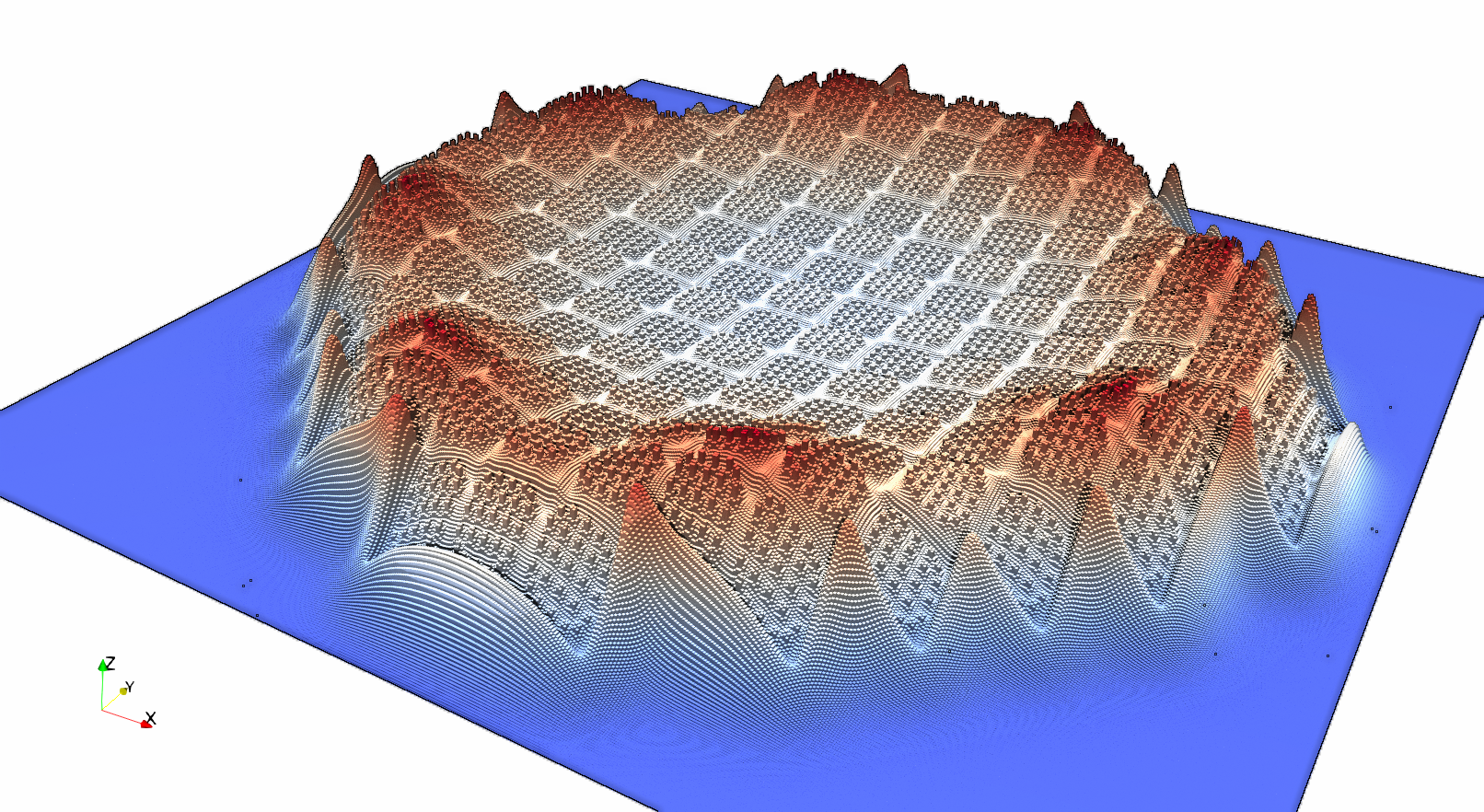}
    \label{fig:PWR900_flux}    
  }
  \caption{Illustration of the 2-Group PWR 900 MW model~\cite{PWR900Tanguy}}
  \label{fig:PWR900}
\end{figure}

These benchmarks correspond to a PWR 900 MW core, and enable $2$, $8$ and $26$ energy groups calculations to be performed. A full
description of these benchmarks is available
in~\cite{PWR900Tanguy}. All benchmarks represent a simplified 3D PWR first
core loaded with $3$ different types of fuel assemblies characterized
by different Uranium-235 enrichment levels (low, medium and highly enriched
uranium). There are no control rods inserted in this core model. Along
the z-axis, the $360$ cm assembly is axially reflected with $30$ cm of
water which results in a total core height of $420$ cm. The $3$ types
of fuel assemblies appear on Figure~\ref{fig:PWR900_materials} where the central
assembly corresponds to the lowest enrichment, while the last row of
fuel assemblies has the highest enrichment to flatten the neutron
flux. Each fuel assembly is a $17 \times 17$ array of fuel pins, with
a lattice pitch of $1.26$~cm that contains $264$ fuel pins and $25$
water holes. The boundary condition associated with this benchmark
problem is a pure leakage condition without any incoming angular flux. The
associated nuclear data, $2$-group, $8$-group and $26$-group
libraries are derived from a fuel assembly heterogeneous transport
calculation performed with the cell code
\dragon~\cite{dragon}. As an example, figure~\ref{fig:PWR900_flux} presents a
visualization of the thermal flux in the central radial plane, as obtained from a
2-group calculation.

Table~\ref{tab:benchmarks_parameters} summarizes the discretization
parameters for the considered benchmarks, where the following
notations are used:
\begin{itemize}
\item $\ngroups$ is the number of energy groups.
\item $N_x$, $N_y$ and $N_z$ define the number of spatial cells along
  the three dimensions of the spatial domain.
\item $\ndir$ is the number of angular directions according to the
  order of angular quadrature in use.
\item $\ndofs$ is the number of degrees of freedoms (DoFs). The
  calculation of DoF numbers consider $3$ DoFs per cell, per energy
  group and per angular direction.
\item Flops is the number of floating point operations required to
  perform a single complete sweep operation, for all energy
  groups. Note that the sweep of a single spatial cell for a single
  angular direction requires $25$ flops (see~\cite{Moustafa2015}).
\item $A_x$, $A_y$ and $A_z$ define the \macrocell sizes along the
  three dimensions. Experimentally, $A_{x,y,z}=16$ was shown to be the most effective choice for the 2-group benchmark.  
\item $\epsilon_{\keff}$ and $\epsilon_{\psi}$ define the thresholds
  used to check the stopping criteria at iteration $n+1$ of the power
  algorithm for the eigenvalue and on the fission source respectively
  as follows:
  \begin{equation}
    \label{eq:stop_keff}
    \frac{|\keff^{n+1}-\keff^n|}{\keff^n}<\epsilon_{\keff},
    \quad
    \frac{||\mathcal{F}\psi^{n+1}-\mathcal{F}\psi^n||}{||\mathcal{F}\psi^n||}<\epsilon_{\psi}.
  \end{equation}
\item  $I_g$ is the fixed number of Gauss-Seidel iterations for the
  multigroup problem.
\end{itemize}
\begin{table}[!htpb]
  \centering
  \begin{tabularx}{\textwidth}{c c c c c c c c c c c }
    \tthickrule
    $\ngroups$&$N_x$&$N_y$&$N_z$&$\ndir$&$\ndofs$&Flops&$A_{x,y,z}$&$\epsilon_{\keff}$&$\epsilon_{\psi}$&$I_g$\\
    & & & & &\footnotesize{$\times 10^{12}$}&\footnotesize{$\times 10^{12}$} & & & &\\
    \hline
    $2$ &$578$ &$578$  &$756$ &$168$ &$0.254$   & $2.12$  & $16$ & $10^{-6}$ &$10^{-5}$&1\\
    $8$ & $578$ & $578$ &$168$ &$80$ &$0.108$   & $0.90$ & $20$ & $10^{-5}$ &$10^{-5}$&5\\
    $26$ & $578$ & $578$ &$140$ &$288$ &$1.051$ & $8.75$  & $20$ & $10^{-5}$ &$10^{-5}$&4\\
    \tthickrule
     \end{tabularx}
 \caption{Description of PWR benchmarks and calculation parameters.}
  \label{tab:benchmarks_parameters}
  \end{table}

Note that the spatial mesh used for the PWR benchmarks is based on a
pin-cell mesh in the $x-y$ plane. Each pin-cell is then subdivided
into $70$ (resp. $84$) slices in the $z$ direction for the $26$-group (resp. $2$-group and $8$-group). The spatial mesh is then further refined by $2 \times 2 \times 2$
for the $8$-group and $26$-group, and by $2 \times 2 \times 9$ for the
$2$-group. The larger spatial mesh for the $2$-group case enables the
study of the strong scalability of our implementation at high core
count.

\bigskip

The experiments have been conducted on computing nodes
(dual Intel Xeon E5-2697v2 processors) of the \athos cluster at EDF.
The theoretical peak performance of each node is 518 (resp. 1036) GFlop/s in double (resp. single) precision ($2\times12$ AVX cores at 2.7 GHz). The following experiments were conducted by launching one MPI process
per computing node and as many threads as available cores; keeping one
core per node for the \parsec communication thread. All experiments
were conducted in single precision. Computation times do not include
setup (reading of cross-section files from the hard disk), but include
all communications and stopping criterion checks. For all the
experiments presented in the following sections, the setup time is
less than a minute.

\subsection{Strong scalability of $2$-group PWR $\keff$ computation}
\label{sec:pwr_calculations}
In this section, we present full-core $\keff$ computations using the
$S_{12}$ $2$-group 3D PWR core model. From a preliminary study with a
single subdomain, we found that the optimal number of $\spn$
iterations is one, for each of the three benchmarks. Therefore, all
the following results are obtained using this value.

\begin{figure}[!htbp]
  \centering
  \subfloat[\pdsa convergence]{
    \includegraphics[width=0.49\linewidth]{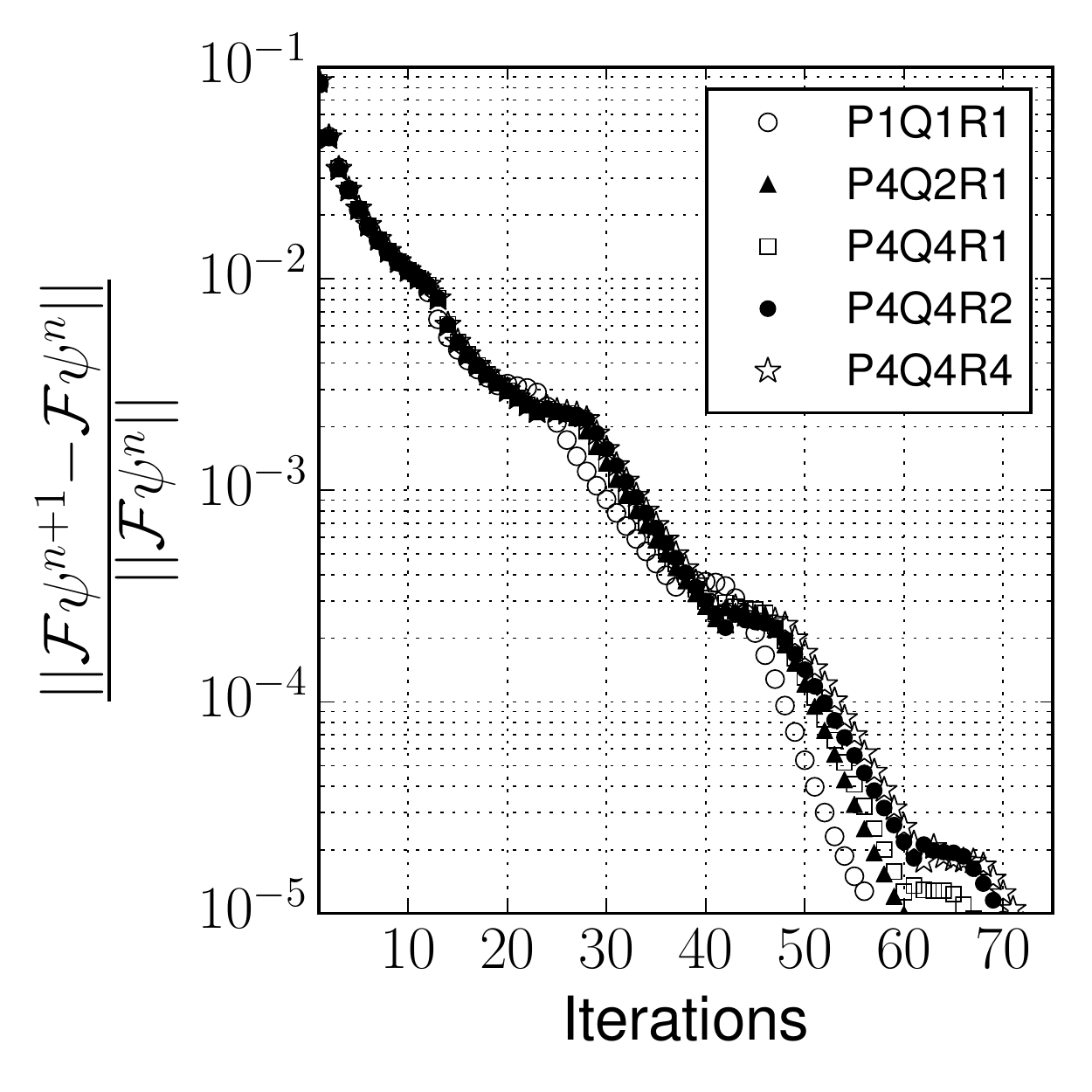}
    \label{fig:pwr2g_convergence}    
  }
  ~
  \subfloat[\pdsa timings]{
    \includegraphics[width=0.48\linewidth]{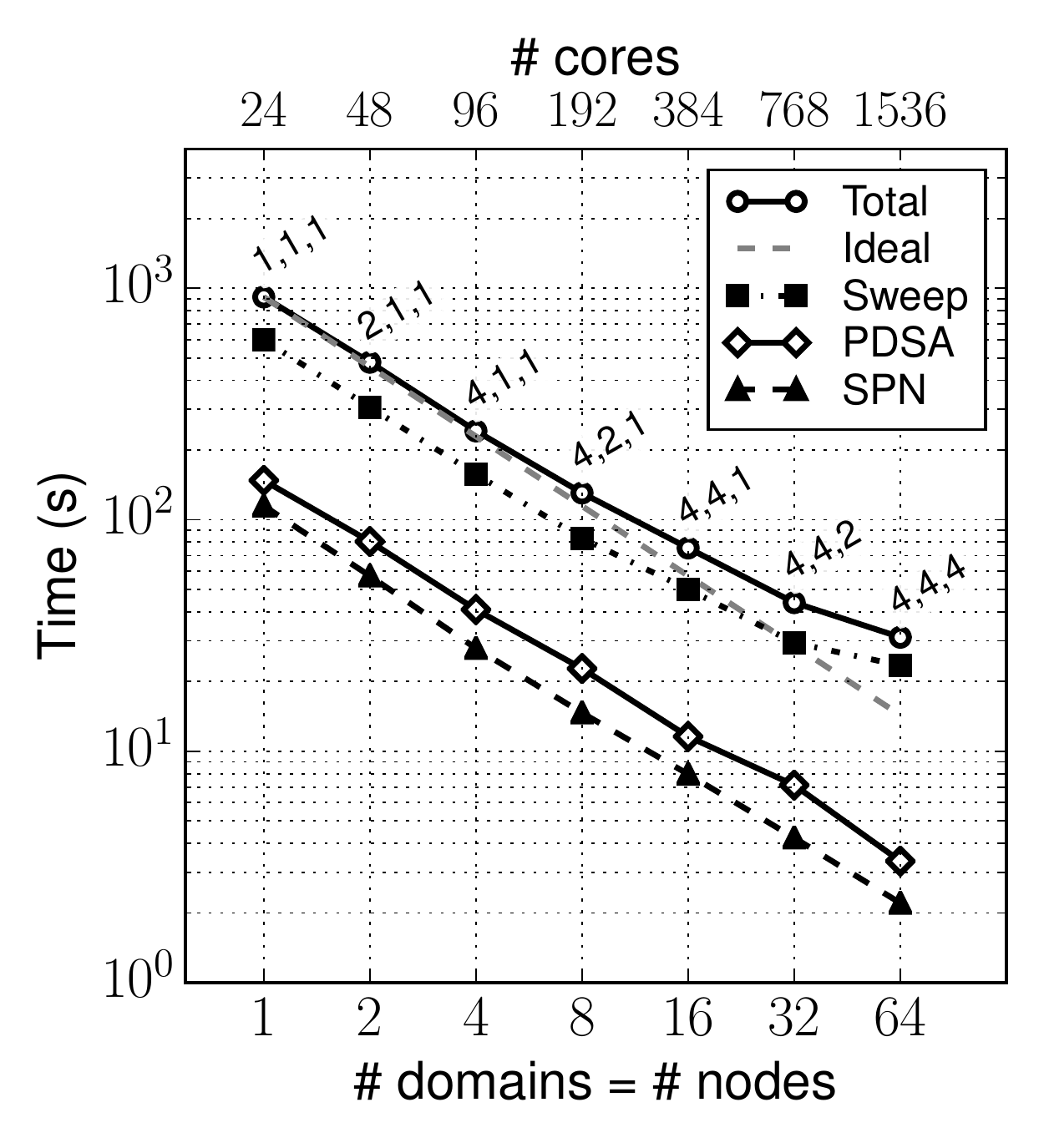}
    \label{fig:pwr2g_Speedup}    
  }
  \caption{Convergence and elapsed CPU time of \domino using the $2$-group PWR
    benchmark for various multi-domain configurations. A $(P,Q,R)$ configuration divides the spatial domain in $P$ (resp. $Q$,$R$) slices in the $X$ (resp. $Y$,$Z$) direction.}
  \label{fig:pwr2g_all}
\end{figure}

Figure~\ref{fig:pwr2g_convergence} compares the convergence of the standard DSA method
with one subdomain $(P1,Q1,R1)$ and those of \pdsa with $4$, $16$, $32$ and $64$ subdomains.
All the computations lead to the same $\keff$ (1.019574) and to the same fluxes.
The outer iteration number increases from $56$ for one subdomain, to $71$ for $64$ subdomains.
This small increase demonstrates that the \pdsa method is a suitable parallel acceleration technique
for PWR core problems. Table~\ref{tab:pwr2g_results} summarizes these DSA and \pdsa results and compares
them to the non-accelerated computation which requires $315$ outer iterations to reach the convergence criterion.
Figure~\ref{fig:pwr2g_Speedup} illustrates \domino's strong scalability. The total computing
time evolution and its main components are displayed. The $\spn$ time, which  corresponds to the time
dedicated to solving all the \pdsa diffusion subproblems in all subdomains, scales perfectly and remains negligible
for all the parallel range. The global parallel efficiency decrease is mainly due to the scalability limitation of the
sweep. These results show that the \pdsa method allows \domino to achieve very good
performance for PWR criticality computations.

%% Accel               & No DSA    & DSA        & PDSA      & PDSA      & PDSA           & PDSA\\

\begin{table}[!htpb]
\begin{tabularx}{\textwidth}{ c l | r | r r r r r}
\tthickrule
($P$,$Q$,$R$)       && $(1,1,1)$ & $(1,1,1)$& $(4,2,1)$ & $(4,4,1)$ & $(4,4,2)$ & $(4,4,4)$\\
Accel               && No DSA    & DSA        & \pdsa    & \pdsa    & \pdsa    & \pdsa\\
 \hline
{$\ncores$}         &&  $24$     & $24$       & $192$     & $384$    &  $768$   & $1536$ \\
 \hline  
 $\nouter$          &&  315      & $56$       &  $59$    & $66$     & $69$     & $71$   \\
 \hline  
  $\tsweep$&(s)      & 3610     & $598.0$    & $83.2$   & $50.0$   & $29.4$   & $23.5$ \\
  $\tspn$  &(s)      &  -        & $114.7$    & $14.6$   & $8.0$    & $4.2$    & $2.2$ \\
  $\taccel$&(s)  &  -        & $147.0$    & $22.8$   & $11.6$   & $7.1$    & $3.4$ \\
 $\ttotal$       &(s)& 4547    & $916.4$    & $130.5$  & $75.5$   & $43.8$   & $31.1$ \\
 \hline  
$\%$ sweep           &&   79   & $65$       & $64$     & $66$     & $67$     & $75$\\
 \tthickrule
\end{tabularx}
\caption{Solution times for a $S_{12}$ $2$-group 3D PWR $\keff$
    computation on the \athos platform.}
  \label{tab:pwr2g_results}
\end{table}

The distributed memory
nodes are efficiently used by this domain decomposition approach. Each multi-core node
parallel potential is efficiently exploited by multi-threaded implementation of
the sweep and mono-domain diffusion solvers. The core SIMD units, handling the third and innermost
parallel level, are efficiently used to simultaneously compute several components of
the angular flux. These three nested levels of parallelism allow \domino to exploit a large
fraction of the computing power of the parallel platform. For this $S_{12}$ (resp. $S_{16}$) 2-group PWR Benchmark
running on 768 cores, the performance of the sweep operation reaches 5.0 (resp. 6.6) TFlop/s  which corresponds
to $15\%$ (resp. $20\%$) of the peak performance of the corresponding 32 nodes.

\subsection{PWR core mode with $8$ and $26$-groups}
Table~\ref{tab:pwr8g_results} presents performance results of 
$S_{8}$ $8$-group and $S_{16}$ $26$-group 3D PWR $\keff$ computations using $64$ computing
nodes of the \athos cluster, partitioned into $(4,4,4)$.
\begin{table}[!htpb]
  \centering
  \begin{tabularx}{\textwidth}{c | Y Y Y Y Y}
    \toprule
    3D PWR &  $\nouter$ & $\tsweep$ (s) & $\tspn$ (s) & $\tcommaccel$ (s) & $\tdomino$ (s) \\
    \midrule
    $S_{8}$ $8$-group & $65$      & $91.53$  & $7.84$   & $0.86$ & $128.85$   \\
    \midrule
    $S_{16}$ $26$-group & $126$     & $2226.42$ & $56.56$ & $147.2$ & $2763.52$  \\
    \bottomrule
  \end{tabularx}
  \caption{Solution times for a $S_{12}$ $8$-group and $S_{16}$ $26$-group 3D PWR $\keff$
    computation on 64 cluster nodes (4,4,4).}
  \label{tab:pwr8g_results}
\end{table}
The convergence on the
 8-group benchmark is reached in $65$ external iterations, and the obtained
eigenvalue is $\keff=1.009408$. This number of external iterations is
similar to what was obtained for a run with a single subdomain. The
total computation time is $128.85$~s of which $91.53$~s comes from the
sweep operation, illustrating that the sweep operation is still
dominant ($71\%$ of the total time).

For the 26-group case, the convergence is reached in $126$ outer iterations, for a global
solver time of $2763.52$~s. The obtained eigenvalue is
$\keff=1.008358$. As in the case with $8$-groups, we did not observe any
increase in the number of external iterations as compared to a run
with a single domain. This is a remarkable result, highlighting the
perfect efficiency of the \pdsa method on representative benchmarks of
our target applications.

\section{Conclusion}
\label{sec:concl}
In this paper, we studied the performance of our
massively parallel approach for solving the neutron transport equation
according to the discrete ordinates method. We first presented our
task-based implementation of the sweep with \parsec, as implemented in
the \domino solver. Then we presented an application of \pdsa, a new
piecewise diffusion acceleration scheme for the scattering iterations. This is required to speed-up
the convergence for strongly diffusive problems. The efficiency of the massively parallel
\pdsa approach has shown to be perfectly effective on different PWR nuclear core criticality computations
where it matches the standard DSA convergence rate.  The Cartesian transport solver \domino,
implementing the \pdsa scheme, exhibits three nested levels of parallelism and exploits a large fraction of the theoretical peak performance of thousands of SIMD computing cores.
As a result, \domino can complete very large and accurate criticality computations
involving more than $10^{12}$ degrees of freedom in less than an hour using 64 super-computer nodes.

This result allows us now to consider future fast and accurate
3D time-dependent transport solutions for diffusive problems.

%\section*{References}
\bibliographystyle{elsarticle-num}
\bibliography{biblio}

\end{document}